\documentclass[a4paper,english,oneside,11pt]{article}
\usepackage{amsmath,amssymb,mathrsfs}
\usepackage{graphicx,epsfig}
\usepackage{color}
\begin{document}

\baselineskip 0.6cm

\begin{titlepage}

\begin{flushright}
\end{flushright}

\vskip 1.0cm

\begin{center}
  {\Large \bf NDA and perturbativity in Higgsless models}

  \vskip 1.0cm

  {\large Michele Papucci}

  \vskip 0.5cm

  {\it Scuola Normale Superiore and INFN, Piazza dei Cavalieri 7,
    I-56126 Pisa, Italy} \\

  \vskip 1.0cm

  \abstract{We analyze unitarity violation in 5D Higgsless theories
  using tree level unitarity bounds on gauge vector bosons
  scatterings. Both elastic and inelastic amplitudes are analytically
  computed and used to estimate the perturbativity cutoff. The results
  are compared with 5D NDA estimates. The inelastic channels are
  essential for a correct determination of the cutoff when light KK
  states are present. In flat space Higgsless models the cutoff is
  always quite low, and can be raised
  at 5 TeV only if one has the first KK state at around 500 GeV, which
  is disfavored by Electroweak data.}

\end{center}
\end{titlepage}

\section{Introduction}

Models where Electroweak Symmetry Breaking (EWSB) is triggered by
boundary conditions for the gauge fields along a compact extra
dimension have been proposed during the last year \cite{Csaki:2003dt,Csaki:2003zu,Barbieri:2003pr}.
In 4D if the Higgs is removed from the Standard Model (SM), the 
scattering amplitudes of the longitudinally polarized vector bosons
grow quadratically with the energy and rapidly violate
unitarity bounds.
Similarly, in these five dimensional models there is no Higgs boson,
but here unitarity of
scattering amplitudes is recovered by the presence of Kaluza Klein
(KK) replicas of gauge
bosons \cite{Csaki:2003dt,SekharChivukula:2001hz}.
Further investigations have shown the similarities between these models
and Technicolor (TC) models in 4D where the Electroweak Symmetry is
dynamically broken by some kind of strong interacting physics \cite{Csaki:2003zu,Barbieri:2003pr,Burdman:2003ya}. In particular, this connection is
transparent in warped models using the AdS/CFT duality. Here the 5D KK states
correspond to resonances of the strong sector of the TC-like dual 4D theory. Moreover, some of the
results retrieved in warped models  using the duality remain also valid in general metrics.
Differently to standard TC models,
in 5D Higgsless theories the first ``resonances'' can appear in an energy range
where the theory is still {\it perturbative}. In this case calculability is clearly
 improved. Moreover early Naive Dimensional Analysis (NDA) \cite{Manohar:1983md}
estimates of the perturbativity range indicated a
cutoff lying in the \mbox{10 TeV} region, parametrically higher than
the one in 4D Higgsless models.

This has mainly motivated the interest in Higgsless theories as a
novel possibility in addressing the EWSB problem.
A deeper analysis has shown that they share with TC models the production of large corrections
to Electroweak Precision observables. The effects are already large in the
pure gauge sector and get worse when matter is
taken into account \cite{Barbieri:2003pr,Peskin:1990zt,Golden:1990ig,Holdom:1990tc,Csaki:2003sh}. Modifications have been introduced to try to reconcile
them with precision data \cite{Cacciapaglia:2004jz}.  
At present it seems impossible to succeed in this task, and
this kind of theories are almost ruled out by past experiments \cite{Barbieri:2004qk}. 

Yet, it remains interesting to analyze how well qualitative expectations are respected
in these models. In fact, the results coming from NDA estimates can also be
useful in a more general context and they are usable in other
situations where a gauge symmetry
is broken by boundary conditions.

Naive dimensional analysis (NDA) is a powerful tool to estimate
coefficients of operators in various dimensions taking
into account 
loops and geometrical factors \cite{Chacko:1999hg}. However, as it is
stressed by its name, NDA is
{\it naive}, since  it does not account for order one
factors. These are sometimes numerically as big as the
``geometrical'' $\pi$ factors meaningfully taken into account. 
In the following we want to test with an explicit
calculation how robust some quantitative conclusions derived by NDA can be. In
particular we are interested in the cutoff of a 5D Higgsless model as
well as in its relation with the one of the
correspondent 4D theory.
The method used in the explicit determination of the cutoff is the
analysis of tree level unitarity in gauge vector bosons scatterings.

In Sect. \ref{sec:NDA} we recall the NDA results of the 5D cutoff of a
Higgsless theory.
In Sect. \ref{sec:scattering} we set up the playground for the explicit
calculation of the $WW \rightarrow WW$ scattering amplitudes, both
elastic and inelastic. In Sect. \ref{sec:elicity} we define the
analysis of tree level unitarity using elicity amplitudes. The results
of the unitarity cutoff are then presented in Sect. \ref{sec:results} and
compared with NDA estimates, while conclusions are drawn in Sect. \ref{sec:conclusions}.

\section{NDA estimation of the cutoff}\label{sec:NDA}

It is well known that a 4D Yang-Mills theory, in
which the gauge symmetry is broken by some kind of strongly interacting
physics, becomes strongly coupled at energies of the order of $4 \pi v
$ where $v$ is the symmetry breaking scale. In terms of $v$ the mass
of the gauge bosons is given, as usual, by $M=g v/\sqrt 2$ where  $g$ is
the coupling constant. This implies
$\Lambda_4 =4 \sqrt 2 \pi M/g $. Thus in the case of the usual 4D SM
without the Higgs boson the cutoff is $\Lambda_4 \simeq 2 TeV$.

We are interested in comparing this result with the 
Higgsless model one. Therefore it is necessary to repeat the NDA
estimate in a 5D Yang-Mills theory broken by boundary conditions on an
orbifold.
 
Let's start recalling the well-known result of  NDA applied to a
5D Yang-Mills theory in a non-compact space
\begin{equation}
  \label{eq:cutoff-5D}
  \Lambda_5=24 \pi^3/g_5^2
\end{equation}

where $g_5$ is the 5D gauge coupling whose dimension is
(mass)$^{1/2}$. This result remains unchanged if we compactify the
fifth dimension on a circle, because the UV cutoff is essentially determined by
local physics and this kind of compactification does not introduce any
modification at short distances. The only requirement is for the
whole picture to make sense, i.e. the new scale $1/R$, introduced by the
compactification, has to be lower than the cutoff (\ref{eq:cutoff-5D}). Things are
different if we compactify the theory on an orbifold, or more
generically on a segment, since here the
physics can be different whether it is probed near one of the fixed points or far
from them. Orbifold projections change UV physics and in general new terms
introduced at the fixed points can change eq. (\ref{eq:cutoff-5D}).

Let us now specialize to Higgsless models. In a five dimensional theory of EWSB without a Higgs, one
 requires only the photon to be a massless particle, while the other
 vector bosons and adjoint scalars (5-th components of vector fields)
 to be massive. We formulate the theory on a segment and these requirements fix the boundary conditions (BC) for all
 the fields. 
It is possible to use gauge freedom
and go in unitary gauge
reabsorbing massive adjoint scalars and leaving only 
massive vector bosons.
At this point, if one does not add any new  interaction
 at the fixed points and repeat NDA analysis in this
 configuration, one finds that eq. (\ref{eq:cutoff-5D}) is still
 valid\footnote{To be more precise, before repeating NDA analysis one should add all the terms
 compatible with the symmetries of the theory, such as boundary
 localized kinetic operators. This will be done later. Here
 we focus on the case in which they are small, i.e. in the large
 radius limit.}. An equivalent procedure consists in  adding boundary localized mass terms to the vector
 bosons. In the limit $m\rightarrow \infty$ one would get the same
 results as imposing BC. In this alternative setup one can
  consider these mass terms as originating by non-linear
 sigma models in the limit $f\rightarrow\infty$
 \cite{Barbieri:2003pr}. Using a Feynmann-'t Hooft gauge where localized Goldstone bosons and
 adjoint scalars are explicitly present, the validity of
 (\ref{eq:cutoff-5D}) could be easily checked. Moreover a direct
 check will be provided by the explicit
 determination of the unitarity cutoff shown below.
With this approach it can also be shown that the absence of any massless
 scalar is not maintained if we add mass terms localized at
 both the boundaries. In fact one can use gauge freedom to
 eliminate Goldstone bosons from one of the non linear sigma models
 localized at the boundaries, but the Goldstones at the
 other boundary will be present in the spectrum, as it is also clear from
 its deconstructed version.

Let us now return to the NDA estimate (\ref{eq:cutoff-5D}). To
  compare it with what NDA says on the cutoff of a   standard 4D TC, one has to
plug in (\ref{eq:cutoff-5D}) the relation between $g_5$ and the
effective 4D coupling $g$, and express the W mass in terms
of $R$.

The relation between $g_5$ and $g$ can be obtained integrating along
the compact dimension the $W$-boson wave-functions
(which are not flat since the $W$'s are massive). To do this one can either use
the cubic or the quartic vertex\footnote{It turns out that the two results are numerically equal
but the position of the $\pi$ factors in the two formulae is completely
different. This is a first warning about the attitude of
  taking ``seriously'' only $\pi$'s while neglecting other  numerical
  factors when doing quantitative estimates with 5D NDA. In the case of a flat extra dimension one gets $
  g_4 = (8 \sqrt 2/3 \pi)(g_5/\sqrt{2 \pi R})$
  from the cubic coupling and eq. (\ref{eq:5d-4d-rel}) from the
  quartic interaction. This can be traced back to the different
  ``KK-number conservation'' properties of the cubic and quartic vertices.}.
Using the 4W vertex for the determination, it is easy to see that in the previous setup one has
\begin{align}
  \label{eq:5d-4d-rel}
  g_4 &= \frac{g_5}{\sqrt{2 \pi R}}\sqrt{\frac{3}{2}}\\
  M &=\frac{1}{2 R}
\end{align}
which gives
\begin{equation}
  \label{eq:5D-cut-new}
  \Lambda_5=\frac{9 \pi}{g \sqrt 2} \Lambda_4
\end{equation}
This analysis suggests that the cutoff of a 5D Higgsless theory is
higher by a factor $k \pi/g$ with respect to the 4D one with the same mass
and coupling. To better understand the origin of this increase it is useful to introduce the
parameter $N\equiv \Lambda_5 R$, which in flat geometries is related to
the number of KK states below the cutoff. One can then rewrite the previous
equations in terms of $N$ and of $\Delta M_{KK}$, the mass splitting
between 2 KK gauge bosons. We get
\begin{align}
  \label{eq:rel-N-Mkk}
  g_5 &= 2 \pi \sqrt{\frac{6 \pi}{N \Delta M_{KK}}} \\
  \Lambda_5 &= N \Delta M_{KK} \\
  \Lambda_4 &=\frac{2}{3}\sqrt N \Delta M_{KK} \\
  \Lambda_5/\Lambda_4 & = \frac{3}{2} \sqrt N \label{eq:rel-5-4-bulk}
\end{align}

From eq. (\ref{eq:rel-5-4-bulk}) we can see that the before-mentioned increment is due to
the presence of the additional KK states below the cutoff. They remedy for
the bad energy behavior of the amplitudes and postpone unitarity
violation. Therefore the effect is proportional to $\sqrt N$.

However this analysis is {\it naive} and order 1 factor in the
coefficients of the previous expressions  can numerically modify the results, potentially
reabsorbing the increment, especially when $N$ is not very large.

Moreover these are the results for the simplest 5D theory, with equally spaced
KK states. Unfortunately it is not phenomenologically interesting,
because the spacing is $1/R \equiv 2 M_W$, too small to be in
accordance with the experiments.

It is therefore necessary to create a bigger separation between the
lowest states and the heavier ones. Furthermore one has to
make the excited states of vector bosons less coupled to the low energy
theory in the hope to reconcile the model with the EWPT. This can be achieved either by progressively increasing the size of
the coefficient of boundary localized kinetic terms for gauge fields
\cite{Barbieri:2003pr} or by using a warped background \cite{Csaki:2003zu}.  
From now on we will follow Ref. \cite{Barbieri:2003pr} and  focus on the case of a
flat background with a brane-localized kinetic operator.

This term introduces a new
parameter in the cutoff relation and in principle can modify or even
lower the cutoff.

NDA still suggests that, apart from ${\cal
  O}(1)$ factors, this is not the case
and that in this kind of setup such a localized kinetic term does not
modify the qualitative behavior of the cutoff.

A way to see this fact would be to repeat NDA in a different gauge as
indicated in \cite{Luty:2003vm}. It is sufficient to see that the UV behavior of
Goldstones and adjoint scalars' propagators is not changed by the localized kinetic term. NDA goes like in
the standard bulk case and
eq. (\ref{eq:cutoff-5D}) remains unchanged.

Thus the main effects of a kinetic term are the change in
the relation between the 4D and the 5D coupling
constant and in the relation between $M_W$ and the 5D
parameters $g$, $g_5$ and $L=\pi R$. The
relation between $\Lambda_5$, $\Lambda_4$ and $N$ are changed consequently.
The relevant equations, when $g^2 L/g_5^2$ is small\footnote{In our
  notation the
  coefficient of the localized kinetic term is $1/g^2$, while the one
  of the bulk term is $1/g_5^2$.}, are
\begin{align}
  & g_4 \simeq g (1-\frac{g^2 L}{4 g_5^2})\\
  & M \simeq \frac{g}{g_5 \sqrt L}(1-\frac{g^2 L}{6 g_5^2})\\
  & \Lambda_4 \simeq \frac{2}{\sqrt 3}\frac{\sqrt N}{\pi}\Delta
  M_{KK}(1+\frac{N g_4^2}{288 \pi^2}) \\ \label{eq:N-def}
  & \Lambda_5/\Lambda_4 \simeq \frac{\sqrt 3 \pi}{2}\sqrt N (1-\frac{
  N g_4^2}{288 \pi^2}) 
\end{align}

A few words of comment are necessary. Eq.
(\ref{eq:cutoff-5D}) gets corrected
when $g^2 L/g_5^2$ become very large, i.e. when higher order
corrections from localized kinetic terms become important.
Trusting $\pi$ factors, NDA suggests that the coefficient of
$\sqrt N$ in the $\Lambda_5/\Lambda_4$ ratio is bigger than the one in
the pure bulk result by a factor $\pi /\sqrt 3$. This means that not only
the cutoff of a 5D Higgsless theory is parametrically higher than the
one in 4D due to the effect of the KK tower, but there is also the
possibility to increase this gain by another ``geometrical'' constant
factor by increasing the separation between the low lying state and
the higher ones \cite{Rattazzi}. Since this result can be of phenomenological
interest but it is retrieved using NDA estimates and is ${\cal O}(1)$, it needs to be
checked whether it may be  canceled by other ${\cal
  O}(1)$ factors not taken into account by NDA. 

\section{The $WW \rightarrow WW$ scattering}\label{sec:scattering}

Let us now begin defining the model used in the analysis before and then we will
describe the calculation of the  $2\rightarrow2$ scattering for $W$ bosons.
The helicity amplitude analysis will be illustrated in the next
section.

We consider a 5D gauge theory compactified on a segment $[0,L]$,
parameterized by the coordinate $y$.  For simplicity we assume the
gauge group to be $SU(2)$, completely broken by boundary conditions at $y=L$. This
  assumption is simple but general enough to apply also to the flat space Higgsless models in the
limit $g' \rightarrow 0$ \cite{Barbieri:2003pr}.  As stated before the gauge
breaking can be achieved by adding by hand  to
the gauge fields a mass term localized at $y=L$.

The relevant Lagrangian is then given by
\begin{equation}
  \label{eq:lagrangian}
  {\cal L}={\cal L}_5 + {\cal L}_0 \delta(y)+{\cal L}_L \delta(y-L)
\end{equation}
where
\begin{align}
  {\cal L}_0 =& - \frac{1}{4 g^2} F_{\mu\nu}^a F^{\mu \nu \ a} \\
  {\cal L}_5 =& - \frac{1}{4 g_5^2} F_{M N}^a F^{M N \ a} \\
  {\cal L}_L =& - m^2 W_{\mu}^a W^{\mu \ a}
\end{align}
where Greek letters indicate 4D coordinates while capital letters the
5D ones, and $a=1,2,3$ is the $SU(2)$ gauge index. Letting $m \rightarrow \infty$ we enforce the
conditions $W_{\mu}^a =0$ at $y=L$ for every $a$. Moreover one can use
the residual 5D gauge freedom to eliminate $W_5^a$, thus working in
unitary gauge.

In order to study tree level unitarity in gauge vector boson
scatterings one way to proceed is to expand all the field in KK modes,
compute the relevant 4D Feynmann graphs and then sum over the
KK towers of intermediate states. This procedure has been followed in
advance in \cite{Csaki:2003dt,Foadi:2003xa} where the elastic $2\rightarrow2$ process has been
calculated and in \cite{Davoudiasl:2003me} where the results of a
numerical evaluation of inelastic amplitudes are presented.  We do not pursue this way of calculation here, instead we
integrate out the degrees of freedom contained in the bulk and we
write down an ``effective'' Lagrangian at $y=0$, i.e. a generating
functional for the $y=0$ fields' amplitudes. This will be the
tool used to calculate the scatterings \cite{Barbieri:2003pr,Luty:2003vm}.  This is motivated by
\emph{economy} and physically by the fact that in the simplest
Higgsless model setup SM matter couples to gauge bosons\footnote{Indeed the SM vector bosons have wave-functions peaked
  at y=0.} at
$y=0$.

Let $g_5=g/M^{1/2}$. $M L$ then parameterizes the ratio between
brane and bulk couplings. The regimes in which the brane dominates
over the bulk and vice-versa are given by $M L \ll 1$, $M L \gg 1$
respectively.
In order to compute the scattering amplitudes we need
the $y=0$ ``effective'' Lagrangian at second order in $g$.  Neglecting
just for a while gauge and spacetime indexes, the vector bosons can be
expanded perturbatively in $g$ as
\begin{equation}
  \label{eq:w-series}
  W=W_0(x; y)+W_1(x;y)+W_2(x;y)+\ldots
\end{equation}
where $W_n$ is of order ${\cal O}(g^n)$ and the $W_n$s satisfy the
conditions
\begin{equation}
  \label{eq:w-conds}
  W_0 (x;y=0) \equiv \bar W(x), \ \ W_n (x;y=0)=0 \ {\rm if} \ n>0
\end{equation}
Then if we indicate with ${\cal L}_5^{(2)}(W,W)$, ${\cal
  L}_5^{(3)}(W,W,W)$, ${\cal L}_5^{(4)}(W,W,W,W)$ the quadratic, cubic
and quartic part of the bulk Lagrangian respectively, the effective
Lagrangian for the bulk contribution can be written as
  \begin{align}
    \label{eq:L-effective}
    {\cal L}_{eff}=\int_0^L dy & \left[ {\cal L}_5^{(2)}(W_0,W_0) +
      {\cal L}_5^{(3)}(W_0,W_0,W_0) + {\cal
        L}_5^{(4)}(W_0,W_0,W_0,W_0) \right. \nonumber \\ & + \left.
      {\cal L}_5^{(2)}(W_1,W_1) + {\cal L}_5^{(3)}(W_{(1},W_0,W_{0)})
    \right]
  \end{align}
  where the other contributions to the perturbative expansion vanish after partial
  integration and/or use of the conditions (\ref{eq:w-conds}), while the
  round parentheses on $W$ indexes mean symmetrization.  Moreover
  the vector boson fields can be decomposed into transverse and
  longitudinal (gauge-like) components $W=W^t+W^l$.  At zero-order
  level they satisfy the 5D homogeneous equation of motion
\begin{align}
  \label{eq:eqmotion}
  (\partial_5^2+p^2)W_0^t &=0 \\
  \partial_5^2 W_0^l&=0
\end{align}
while at first order level they satisfy the inhomogeneous version of
the previous equations where the r.h.s. is $ 1/2 (\delta {\cal L}_5^{(3)}/\delta
W)(W_0,W_0)$.  At this point one can see that the two terms
in the second line of (\ref{eq:L-effective}) are not independent among
each others because
the equations of motion enforce the relation
\begin{equation}
  \label{eq:lagr-relation}
  \int_0^L dy  \ {\cal L}_5^{(2)}(W_1,W_1)=
-\frac{1}{2} \int_0^L dy \ {\cal L}_5^{(3)}(W_{(1},W_0,W_{0)})
\end{equation}
Before going on a few comments on the physical meaning of
(\ref{eq:L-effective}) are worthly noticing.  The first term  describes the correction to the vector boson propagator because
 the W bosons can leave the $y=0$ brane and propagate in
the bulk. The second and third terms include the corrections to the
interaction vertices because the interaction point can be far from the brane.
The last two terms, instead, describe an effective 4-point interaction
originating from a tree level 2-2 scattering mediated by a vector boson exchange
taking place entirely in the bulk. This can be clearly seen in
eq. (\ref{eq:bulk-scat}) below from
the explicit form of the effective Lagrangian.

Going to momentum space only for the first four coordinates and solving the
equations of motion for the vector bosons one finds
\begin{align}
  \label{eq:sol-eqmotion}
  W^t_0 (k;y) & = \bar W^t(k) \frac{\sin
    \left(\sqrt{k^2}(L-y)\right)}{\sin
    \left(\sqrt{k^2}L\right)} \\
  W^l_0 (k;y) &= \bar W^l(k) \frac{L-y}{L} \\
  W^t_1 (k;y) &= -\bar W^t(k)\sin \left(\sqrt{k^2}(L-y)\right)
  \int_0^y ds \frac{1}{\sin^2 \left(\sqrt{k^2}(L-s)\right)}\int_s^L dt
  \sin
  \left(\sqrt{k^2}(L-t)\right) \frac{\delta {\cal L}_5^{(3)}}{\delta W^t}(k;t) \\
  W^l_1 (k;y) &= -\bar W^t(k)(L-y) \int_0^y ds
  \frac{1}{(L-s)^2}\int_s^L dt (L-t) \frac{\delta {\cal L}_5^{(3)}}{\delta
    W^l}(k;t)
\end{align}
where $(\delta {\cal L}_5^{(3)}/\delta W) (k;t) \equiv (\delta {\cal
  L}_5^{(3)}/\delta W) (W_0(p;t),W_0(q;t))$ and momentum conservation
  $p+q+k=0$ is enforced.
Using these formulae the last two terms in the effective Lagrangian
can be rewritten as
\begin{equation}
  \label{eq:bulk-scat}
\int_0^L dy \left(\frac{\sin
  \left(\sqrt{k^2}L\right)}{\sin
  \left(\sqrt{k^2}(L-y)\right)}\right)^2\left[\int_y^L ds \frac{\delta {\cal
  L}_5^{(3)}}{\delta W^t}\left(k;s\right)\right]^2 +\int_0^L dy \frac{L^2}{(L-y)^2}\left[\int_y^L ds \frac{\delta {\cal
  L}_5^{(3)}}{\delta W^l}\left(k;s\right)\right]^2 
\end{equation}
which clearly describes a 2-2 bulk scattering mediated by either a
longitudinal vector boson or a transverse one.  The propagators for
the transverse and longitudinal components are:
\begin{align}
  \label{eq:propagators}
 \langle \bar W^{t,a}_{\mu} \bar W^{t,b}_{\nu} \rangle (k) &=-i\delta^{ab}
 \left(g_{\mu\nu}-\frac{k_{\mu}k_{\nu}}{k^2}\right)\frac{1}{k(k-M
 \cot(k L))}  \\  
 \langle \bar W^{l,a}_{\mu} \bar W^{l,b}_{\nu} \rangle (k) &=i\delta^{ab}
 \frac{k_{\mu}k_{\nu}}{k^2}\frac{L}{M}  \\  
\end{align}

Since the W fields in the effective Lagrangian (\ref{eq:L-effective})
are a superposition of Kaluza Klein fields of different masses, in
order to calculate scattering amplitudes among mass eigenstates one
has to apply the standard LSZ procedure on external legs multiplying
 each of them by $Z^{1/2}$. The last ingredient needed is then the pole
residue of the zero-order vector boson propagator calculated at
$p^2=m^2_i$, the mass of the i-th KK mode. This can be easily
retrieved from eq. (\ref{eq:propagators}). For a KK state of mass $m$
it is
\begin{equation}
  \label{eq:pole}
  Z^{1/2}=\sqrt{\frac{2}{1+M L \csc(m L)}}
\end{equation}

Having computed the relevant Lagrangian ${\cal L}={\cal L}_0+{\cal L}_{eff}$ we
are able to calculate the $WW\rightarrow WW$ polarized scattering
amplitudes in a straightforward way. In particular all the
contractions between polarization vectors and external momenta work
like in 4D.

\section{Elicity amplitudes analysis of tree level unitarity}\label{sec:elicity}

In order to investigate tree level unitarity one can decompose
scattering amplitudes into elicity amplitudes of definite total
angular momentum. In the following we will focus on the $J=0$
amplitude which, for massive spin 1
particles, is a 3 by 3 matrix in elicity space. It involves
polarized scatterings of vector bosons of same helicity\footnote{Differently from the commonly
  used convention to define elicity amplitudes with all outgoing legs, a
convention closer to our specific situation is used here, in which the first two
lines are incoming while the second two are
outgoing. In this way we will write $(+,+,-,-)$ for the usual
$(+,+,+,+)$ amplitude.}, i.e. $(0,0)$,
$(+,+)$ and $(-,-)$.

It is easily seen that  a general inelastic channel  consists of
five different independent amplitudes, because parity and
time-reversal constrain the 3 by 3 matrix. Usually in a 4D Higgsless Yang-Mills
theory the $(0,0,0,0)$ scattering
amplitude grows linearly with $s$ and dominates upon the other
polarized channels which
are therefore neglected. In this way one can restrict the analysis to a
single entry in elicity space. In the 5D case, this
amplitude is better-behaved and grows logarithmically with $s$ \cite{Csaki:2003dt,SekharChivukula:2001hz}.

One could suspect that the other four elicity amplitudes could not be
neglected anymore. 
However off diagonal amplitudes are smaller than the
elicity-conserving ones. 
This is quite obvious for the $(+,+,-,-)$ amplitude
which is  well-known to be equal to 0 in the massless limit because of
angular momentum conservation. The suppression can be easily derived also for the $(0,0,\pm,\pm)$
amplitudes taking the high energy limit (in this case the
role of the KK tower is crucial, as it is crucial the role of the
Higgs boson exchange in the SM case).

\begin{figure}[ht]
  \centering
\includegraphics[width=.5\textwidth]{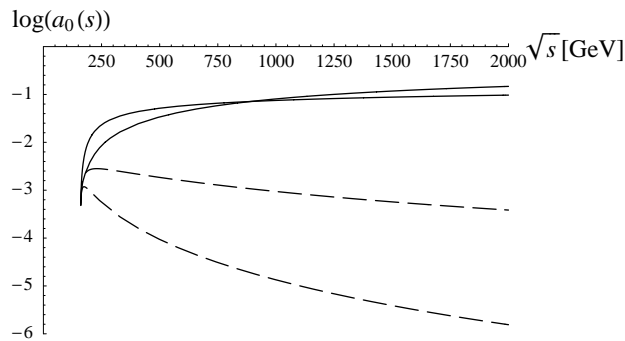}
  \caption{Elastic elicity amplitudes. Solid lines: $(0,0,0,0)$ and $(+,+,+,+)$. Dashed lines: $(0,0,+,+)$ and $(+,+,-,-)$.}
  \label{fig:elicity}
\end{figure}

This fact keeps the $J=0$ matrix nearly
diagonal\footnote{This has been checked for all the
  cases shown below.}, letting us to concentrate to the
$(0,0)\rightarrow (0,0)$ matrix element, which is slightly bigger, but
of the same order, of the other ``elicity-conserving'' amplitudes.
To get a numerical feeling of this statement, Fig. \ref{fig:elicity}
shows the elastic elicity amplitudes as a function of $\sqrt s$.

As to the isospin indices, the $T_3=0$ channel will be
analyzed. Indeed it is the fastest growing one because it involves both the $W^+ W^-$ and $W^0
W^0$ initial and final states. This 2 by 2 matrix can be easily
diagonalized and the biggest  eigenvalue, the $T=0$ one, will be taken
in the following
discussion on unitarity.  

Having get rid of
helicity, angular momentum and isospin indices with the previous
considerations, an infinite matrix remains whose elements are the
scattering amplitudes of different KK modes.

Let us define the transition matrix $T$ for $J=0$ amplitudes in such a way
that its elements are
\begin{equation}
  \label{eq:T-def}
  t_{\alpha \beta}=\frac{1}{32 \pi}\int_{-\pi}^{\pi}d\theta {\cal
  M}_{\alpha\rightarrow \beta}(s;\cos \theta)
\end{equation}
where $\alpha$ and $\beta$ are initial and final states labeled by a
couple of KK indices. 
Unitarity constraints can then be recasted in matrix form as \cite{DeCurtis:2003zt}
\begin{equation}
  \label{eq:unitarity}
T -T^{\dagger}=-2\, i \ T \, \Phi \, T^{\dagger} 
\end{equation}
where $\Phi$ is the (diagonal) matrix of phase
space contributions
\begin{equation}
  \label{eq:phase-space}
  \Phi_{\alpha}=\frac{2 |p_{\alpha}|}{\sqrt s}=\sqrt{1+\frac{(m_n^2-m_k^2)^2}{s^2}-2\frac{m_n^2+m_k^2}{s}}
\end{equation}
where $\alpha=(n,k)$, $s$ is the squared C.M. energy and $| p_{\alpha}|$ is the
absolute value of 3-momentum in C.M. frame of the state $\alpha$.

As a reference let us recall that when this matrix collapses to a single
element $t$ one can derive the usual bound
\begin{equation}
  \label{eq:single-bound}
  | t |< 1
\end{equation}
In principle one should diagonalize the matrix equation in
(\ref{eq:unitarity}) and apply the condition (\ref{eq:single-bound}) to
every eigenvalue.  It is however cumbersome to diagonalize the full
matrix. A  somewhat weaker bound can be derived by focusing on one
single element of the l.h.s., i.e. on the row of $T$ with two
low-lying KK modes as initial state. These  are to be identified
with SM vector bosons.  Unitarity then implies
\begin{equation}
  \label{eq:row-bound}
  \Phi_{0,0} \, |t_{0,0}|+ \frac{1}{|t_{0,0}|} \sum_{(n,m)\neq(0,0)}
  \Phi_{n,m} \, |t_{n,m}|^2<1
\end{equation}
where now $t_{n,m}$ represents the $(0,0)\rightarrow(n,m)$ channel and
the sum is performed over all accessible states for a given
C.M. energy $\sqrt s$. 


\section{Tree-level unitarity and NDA comparison}\label{sec:results}

In this section the scale of the unitarity cutoff is determined by taking
into account both
elastic and inelastic channels. Since all partial wave amplitudes
singularly considered grow logarithmically with $\sqrt s$
\cite{Csaki:2003dt,SekharChivukula:2001hz}, the previous
estimates of the cutoff based on the elastic amplitude alone
\cite{Foadi:2003xa} are in general not
reliable when the excited KK states are not very heavy. 
In fact it is well known that in extra-dimensional models it is the multiplicity of
states, i.e. the increasing number of open channels, that leads to
unitarity violation and not the growing with the energy of a single
amplitude.
  
In principle, since the kinetic term at $y=0$ completely breaks 
5-momentum and  every KK state can interact with all the others, one
can estimate to have ${\cal
  O}(n^2)$ open channels at C.M. energies of the
order of the mass of the n-th Kaluza-Klein. If all the open channels had amplitudes of
the same order one could expect a quadratic growth with the energy of
at least one of the eigenvalues of $T$, hence a dependence of
$\Lambda_5$ on $N_{KK}^2$ which strongly conflicts with NDA. Clearly this is not the case.

The reason is that all the channels which violates momentum along the 5th coordinate are
suppressed. 

In fact, all the transition amplitudes associated to them have a suppression
factor proportional to the sum of the final particles masses,
hence to the sum of the KK indices. 
Restricting the analysis to the 5th momentum-conserving amplitudes only,
$(0,0)\rightarrow (n,n\pm1)$ and $(0,0)\rightarrow (n,n)$, one obtains a transition matrix whose
dimension is ${\cal
  O}(n) \times {\cal O}(n)$. Its largest eigenvalue then scales at
most linearly with the number of accessible KK states, in agreement
with NDA.
Taking into account all the 5th
momentum violating amplitudes, because of their suppression, gives an
${\cal O} (1)$ correction but does not modify qualitatively the energy behavior.

\begin{figure}[t]
  \centering
\includegraphics[width=.7\textwidth]{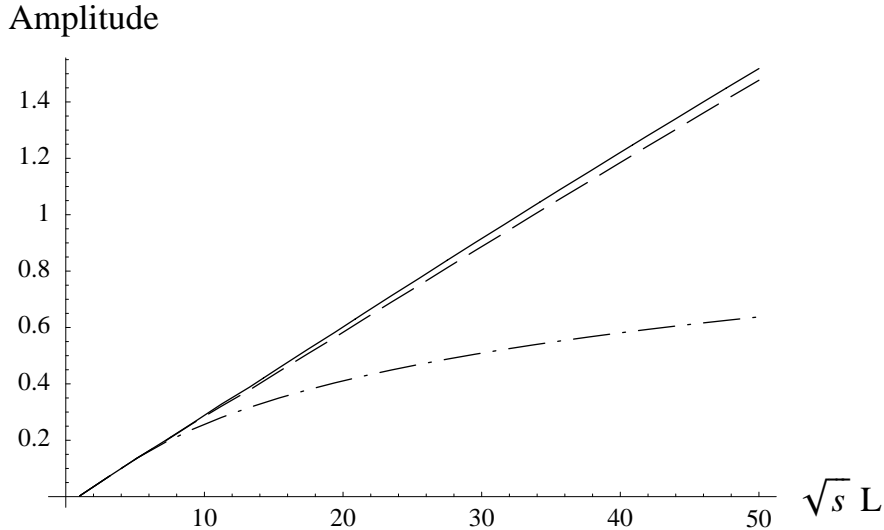}
\caption{Comparison between l.h.s. of eq. (\ref{eq:row-bound}) calculated
  using all the scattering amplitudes (solid line), the
  5th momentum-conserving ones (dashed line) and the elastic
  amplitude alone (dashed-dotted line) for $\sqrt{M L}=0.3$.}
  \label{fig:violation-p}
\end{figure}

As a numerical example of the above statement 
Fig. \ref{fig:violation-p} shows the l.h.s. of (\ref{eq:row-bound})
where either all the amplitudes or only the
$(0,0)\rightarrow(n,n)$ are included in the sum. In the same Figure it is also
plotted the elastic amplitude alone, showing explicitly that inelastic
contributions cannot be neglected.
Having clarified this point, all the
channels will be taken into account in the following computations.

\begin{figure}[t]
  \centering
\includegraphics[width=\textwidth]{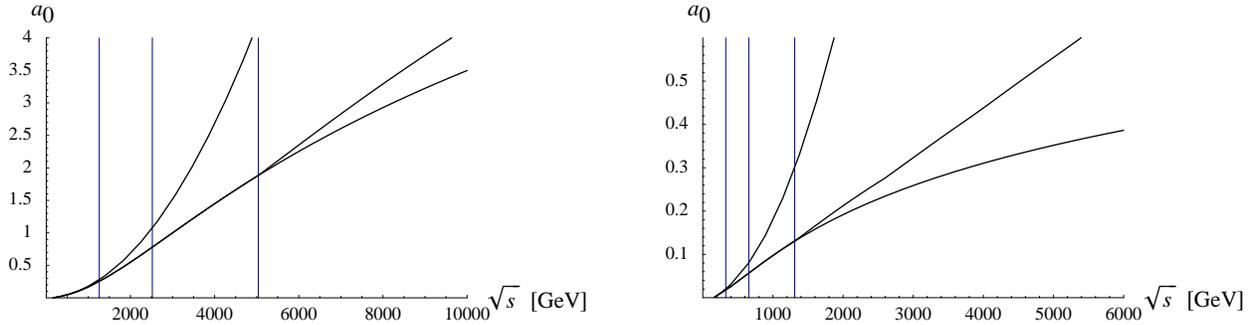}
\caption{Comparison between 4D, 5D elastic and the l.h.s. of
  eq. (\ref{eq:row-bound}) as a function of $\sqrt s$. Vertical lines
  are for $\sqrt s =M_{W'}/2,M_{W'},2 M_{W'}$. The two figures are
  plotted for $\sqrt{ML}=0.1,0.4$ respectively.}
  \label{fig:log-compare}
\end{figure}

It is interesting to compare the 4D amplitude with the 5D elastic and
inelastic ones. This is shown in Fig. \ref{fig:log-compare} for two
different values of $ML$. 
We can see that the amplitudes in the Higgsless model follow the
energy behavior of the 4D one up to energies of the order of $M_{W'}/2$, where $W'$ is the first
excited state. After that the presence
of the KK modes modify the quadratic energy dependence into a
logarithm. It is also clear that the elastic amplitudes gives the
correct behavior up to $2 M_{W'}$ where the channel into
$WW\rightarrow W'W'$ is
open.
In this region the energy dependence is almost linear since the
$\log(\sqrt s/M_{W'})$ is small. On the contrary, when the logarithm
becomes large, other channels are opened and the elastic amplitude
cannot be trusted anymore. 

If the KK states are quite heavy then the amplitude becomes large in
the region where $\sqrt s \lesssim M_{W'}$ and it likely saturates the
bound before inelastic channels are open. In this case the elastic
amplitude suffices in giving a correct estimation of the
cutoff. This is also the region where $N_{KK}$ is very small and 5D
  NDA fails. The cutoff of this 5D Higgsless theory is not too far
from the one of a Higgsless SM. Moreover the effective field theory description, which is
an expansion series in $1/N_{KK}$, is becoming
meaningless when $N_{KK}$ is small. Conversely, if the KK states are light then a full
inelastic computation is required. However Figs. \ref{fig:violation-p}
and \ref{fig:log-compare} suggest that a behavior linear in energy
is early
attained and a linear extrapolation using the angular coefficient
computed from the elastic amplitude in the range $M_{W'} \lesssim \sqrt s
\lesssim 2 M_{W'}$ can be a good approximation of the full result.

Let us now return to the comparison of the scattering amplitudes with
NDA estimates.

From the perturbative unitarity analysis one can define $\bar \Lambda_5$ as the energy scale where the
bound (\ref{eq:row-bound}) saturates. Therefore
$\bar \Lambda_5$ can be determined for a given set of values of $M$, $L$ and
$g$.

One can compare it with $\Lambda_5$ from NDA and check directly both
the $24 \pi^3$ factor in (\ref{eq:cutoff-5D}) and the dependence of
$\Lambda_5/\Lambda_4$ on $\sqrt N$ on (\ref{eq:rel-5-4-bulk}) and (\ref{eq:N-def}).

To check (\ref{eq:cutoff-5D})  it is convenient to adimensionalize both sides of
the equality using $L$.
Since the scattering amplitudes are proportional to $g_5^2/L$, having defined
$\bar N=\bar \Lambda_5 /\Delta M_{KK}$, the unitarity bound reads
\begin{equation}
  \label{eq:N-rel}
  g_5^2/L f(\bar N,M L)=1
\end{equation}
Inverting this relation, one can easily plot $\bar N$ versus $L/g_5^2$ for different values
of the localization parameter $\sqrt {ML}=g \sqrt L/g_5$.

According to NDA a linear relation  with a coefficient
$12 \pi^2$ should be obtained if we identify $N$ with $\bar
N$. However,  as long as eq. (\ref{eq:N-rel}) is derived from an upper bound on
$\Lambda_5$, then $\bar N \geq N$ and a smaller coefficient suffices.

\begin{figure}[htb]
  \centering
\includegraphics[width=.7\textwidth]{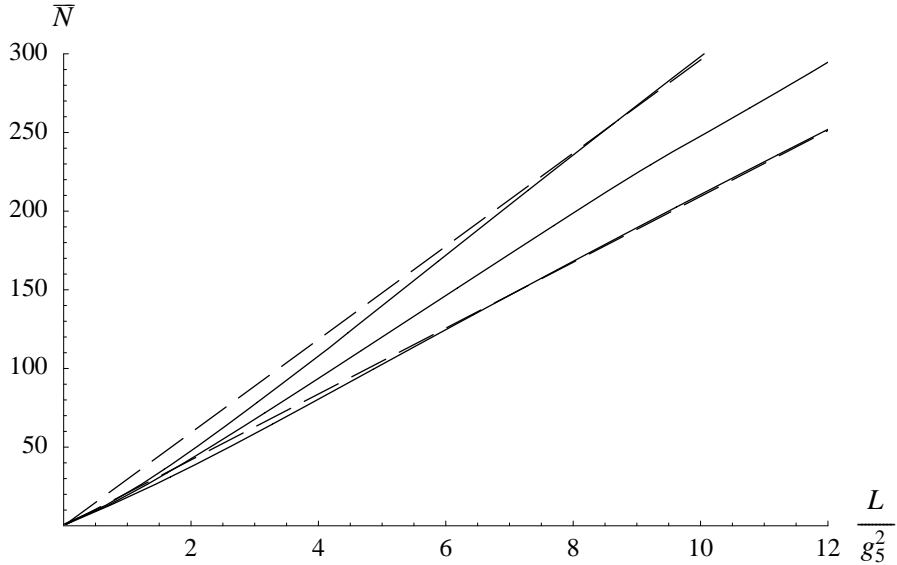}
\caption{Solid lines: $\bar N$ from eq. (\ref{eq:N-rel}) as a function
  of $L/g_5^2$ for $\sqrt{M L}=0.3,1.75,15$. Dashed lines: the
  functions $3 \pi^2 L/\sqrt 2 g_5^2$ and $3 \pi^2 L/g_5^2$.}
  \label{fig:N-funct}
\end{figure}

Fig. \ref{fig:N-funct} shows $\bar N$ as a function
  of $L/g_5^2$ for $\sqrt{M L}=0.3,1.75,15$ compared with two linear
  functions of $\bar N$
  with different coefficients (dashed lines).
As a confirmation of 5D NDA, not only the linear behavior is respected but also the
  coefficients show the presence of the right power of $\pi$. The difference is  only  a
  numerical factor, ranging from $4$ to $4/\sqrt2$ as $ML$ changes from $ML\gg 1$ to $ML \ll 1$.

It is now possible to check the NDA result for
  $\Lambda_5/\Lambda_4$. Here things are simpler. Any overestimation\footnote{Such
  an overestimation is certainly present since stronger bounds can be
  formulated. If we restrict to the case of a $1\times1$
  matrix it is well known that a stronger bound is $Re
  |t|<1/2$. Moreover Figure \ref{fig:N-funct} suggests that the main
  source of the overestimation comes as a multiplicative constant.} in
  the definition of $\bar \Lambda$, if in the form of a multiplicative
  factor, equally affects both $\bar \Lambda_5$ and
$\bar \Lambda_4$ and cancels out. Since the previous result suggested the
  presence of such multiplicative factor, we should get 
  a sufficiently reliable result for $\Lambda_5/\Lambda_4$.
\begin{figure}[ht]
  \centering
\includegraphics[width=.7\textwidth]{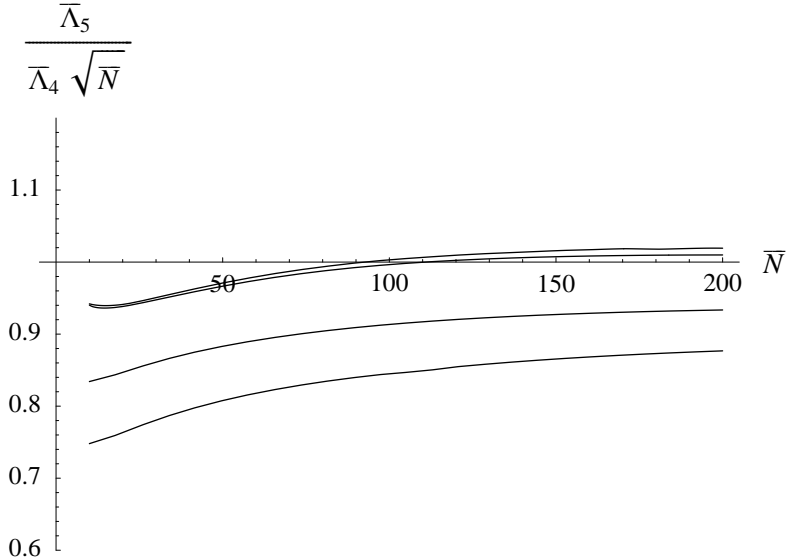}
\caption{$\Lambda_5/\Lambda_4 \sqrt{N_{KK}}$ as a function of
  $N_{KK}$ for $\sqrt{M L}=0.05,0.3,1.75,15$.}
  \label{fig:ratio}
\end{figure}

 Here we are interested in the $\sqrt N$ coefficient. Hence we
 focus on 
 the ratio $\bar \Lambda_5/\bar \Lambda_4 \sqrt {\bar N}$, where $\bar
 \Lambda_5$ and $\bar N$ are defined as before and $\bar \Lambda_4$ is
 the unitarity cutoff of the correspondent 4D gauge theory, retrieved using
 the bound (\ref{eq:single-bound}). Fig. \ref{fig:ratio} shows the plot of this quantity versus $\bar N$.

 One can easily see that the $\sqrt N$ dependence of the cutoff ratio
 is quite good and the that coefficient in eq. (\ref{eq:N-def}) is
 approximately 1 for low values of $M L$ and decreases for higher
 values of the localization parameter (less localization). This confirms the
 qualitative NDA result that there is an increment in the ratio between the 5D and 4D cutoffs
 when a large localized kinetic term is added.  It also shows that the
 numerical values of the $\sqrt N$
 coefficients in both of the cases are different, lower than the NDA
 estimates.
In particular the increase by $\pi /\sqrt 3$, as suggested by NDA, is
 reduced to $1.15$ which is numerically irrelevant. 

\begin{figure}[htb]
  \centering
\includegraphics[width=.7\textwidth]{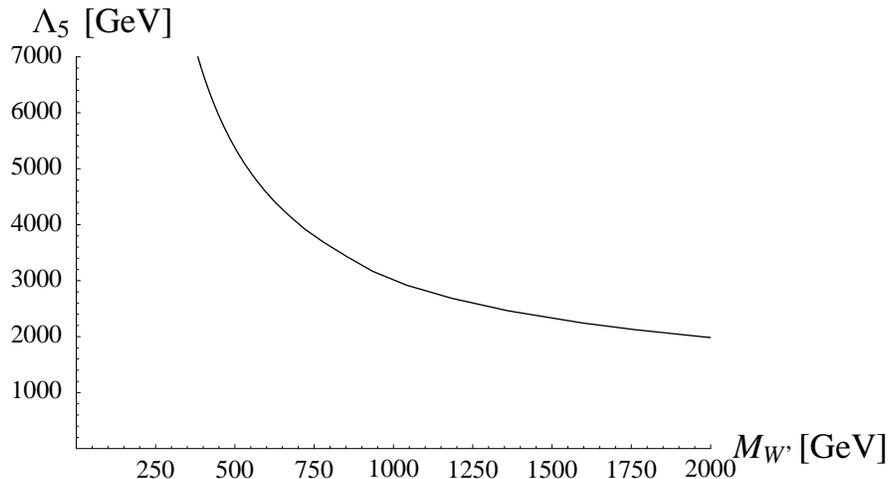}
  \label{fig:SM-cutoff}
\caption{Cutoff of a flat space Higgsless theory as a function of the
  mass of the first excited vector boson.}
\end{figure}

As a final result one can determine the cutoff for a ``realistic''
Higgsless theory setting the low lying state at $m_W = 80 \, GeV$ and
relating the effective coupling $g_{(0,0,0)}$ between three W bosons
to the SM value $g_{SM}$. An additional parameter, $M
L$, is left out and can be determined by the position of the first excited state.
Fig. \ref{fig:SM-cutoff} shows the results for $\Lambda_5$ as a
function of $m_{W'}$. One can see that the cutoff is always less than
about 5 TeV for $m_{W'}> 500 \, GeV$. This result rigorously applies only to the Higgsless
``flat'' models where the introduction of $g' \neq 0$ does not modify the
results. However in \cite{Davoudiasl:2004pw} it is found a numerically similar
result also for warped models, although using a stronger bound $Re \, a_0 <1/2$.
This indicates a very low cutoff for those parameters space regions
where 5D Higgsless theories are not in disagreement with Electroweak
Precision Tests (EWPT) data. In fact EWPT constrain the 
first resonances exchanged in the neutral channel to be quite heavy,
and this lowers the perturbative cutoff. Higher order
operators parameterizing UV physics are weighted by powers of
$1/N=1/\Lambda R$. Since $N$ turns out to be ${\cal O}(1)$, the price to be
paid to be compatible to EWPT is that the same electroweak observables are
affected by uncalculable UV physics with potentially large
corrections. The situation is not very different from the one in 4D
TC, whose calculability these
models aimed to improve.

\section{Conclusions}\label{sec:conclusions}

In this work we have computed both elastic and inelastic partial waves
for 4W scatterings in a 5D Yang-Mills theory broken by boundary
conditions.
These results allowed us to study how reliable NDA estimates are in a
5D Higgsless theory. In particular, we have explicitly checked the
$\sqrt N$ dependence of the ratio $\Lambda_5/\Lambda_4$ between a 5D
Higgsless theory and the correspondent 4D one. We have also
investigated the enhancement of the coefficient of this relation when
a big localized kinetic term is added to the theory. It was found
that this increase, suggested by NDA, is reabsorbed by ${\cal O}(1)$
factors which NDA can't control. Finally we have given a better
determination of $\Lambda_5$ in 5D Higgsless flat models finding a
very low value. $\Lambda_5$ can be raised up to about $5\,  TeV$ only
if $M_{W_3'}\sim 500\, GeV$, which is  however excluded by EWPT. In those parameter space regions where the models are
still compatible with EWPT data the cutoff is $2-3\, TeV$, giving $N
\sim 2$.

\section*{Acknowledgments}

I thank Riccardo Rattazzi for invaluable discussions and
clarifications. This work was supported in part by MIUR and
by the EU under TMR contract
HPRN-CT-2000-00148.

\end{document}